\documentstyle[preprint,aps]{revtex}
\input epsf.tex
\newcommand{\bc}{\begin{center}}
\newcommand{\ec}{\end{center}}
\newcommand{\beqn}{\begin{equation}}
\newcommand{\eeqn}{\end{equation}}
\newcommand{\barr}{\begin{eqnarray}}
\newcommand{\earr}{\end{eqnarray}}

\def\del{\partial}
\def\etal {{\it et al}. }
\def\eg {{\it e.g}. }
\def\ie {{\it i.e}. }

\def\tr {\mbox{tr}}

\def\sin {\mbox{sin}}
\def\PL #1 #2 #3 {Phys. Lett.~{\bf#1} (#2) #3}
\def\NP #1 #2 #3 {Nucl. Phys.~{\bf#1} (#2) #3}
\def\NPP #1 #2 #3 {Nucl. Phys.~{\bf B} (Proc.~Suppl.)~{\bf#1} (#2) #3}
\def\ZP #1 #2 #3 {Z.~Phys.~{\bf#1} (#2) #3}
\def\PR #1 #2 #3 {Phys. Rev.~{\bf#1} (#2) #3}
\def\PP #1 #2 #3 {Phys. Rep.~{\bf#1} (#2) #3}
\def\PRL #1 #2 #3 {Phys. Rev.~Lett.~{\bf#1} (#2) #3}
\def\PTP #1 #2 #3 {Prog. Theor.~Phys.~{\bf#1} (#2) #3}
\def\MPL #1 #2 #3 {Mod. Phys.~Lett.~{\bf#1} (#2) #3}
\def\RMP #1 #2 #3 {Rev. Mod.~Phys.~{\bf#1} (#2) #3}
\def\IJM #1 #2 #3 {Int. J.~Mod.~Phys.~{\bf#1} (#2) #3}
\begin{document}
\draft
\preprint{November 1998}
\title{
Topological Aspect of 
Abelian Projected SU(2) \\
Lattice Gauge Theory
}
\author{
Shoichi~SASAKI$^{\;a)}$\footnote[3]
{E-mail address:~ssasaki@bnl.gov}
 and Osamu~MIYAMURA$^{\;b)}$}
\address{a) RIKEN BNL Research Center, Brookhaven National Laboratory, 
Upton, NY 11973, USA}
\address{b) Department of Physics, Hiroshima University, 
Higashi-Hiroshima 739-0046, Japan\\
}
\date{Nov 23, 1998}
\maketitle
\renewcommand{\thefootnote}{\fnsymbol{footnote}}
\newlength{\minitwocolumn}
\setlength{\minitwocolumn}{7.0cm}
\setlength{\columnsep}{0.8cm}
\renewcommand{\thefigure}{\arabic{figure}} 
\newcommand{\fcaption}[1]{\refstepcounter{figure} Fig.~\thefigure  #1}
\begin{abstract}
We show that the hypothesis of abelian dominance allows QCD-monopoles to preserve
the topological feature of the QCD vacuum within SU(2) lattice gauge theory.
An analytical study is made to find the relationship between the topological 
charge and QCD-monopoles in the lattice formulation. 
The topological charge is explicitly represented 
in terms of the monopole current and the abelian component of gauge fields 
in the abelian dominated system.
We numerically examine the relation and demonstrate the abelian dominance 
in the topological structure by using Monte Carlo simulation.  
\end{abstract}

\vspace{0.5cm}

\vfill\eject
\section{Introduction}
It is important to understand topological aspects of QCD
to interpret several basic properties of the vacuum structure.
Two distinct pictures of the QCD vacuum, which are based on the 
presence of topological objects; instanton and QCD-monopole, 
are now widely accepted. The main reason is that they give nice descriptions of 
several non-perturbative features, \eg chiral symmetry breaking and color 
confinement. As is well known, ${\rm SU}(N_{c})$ gauge theory has classical 
and non-trivial gauge 
configurations (instantons) satisfying the field equation in the 4-dimensional 
Euclidean space ${\bf R}^4$ \cite{Instanton}. The topological charge, 
which corresponds to the non-trivial homotopy group 
$\pi_{3}({\rm SU}(N_{c}))=Z_{\infty}$, is assigned to instanton 
configurations \cite{Instanton}.
It has been established that such a topological feature plays an essential role 
on the resolution of the ${\rm U}_{\rm A}(1)$ problem \cite{Hooft1}. 
Furthermore, the instanton liquid characterized 
by a random ensemble of instanton and anti-instanton configurations 
succeeds in explaining several properties of light hadrons, \eg
spontaneous chiral-symmetry breaking (S$\chi$SB) \cite{Instanton}.

Another picture of the QCD vacuum is motivated from the stimulating idea of 
abelian gauge fixing, which was proposed by 't Hooft \cite{Hooft2}
and also independently by Ezawa-Iwazaki \cite{Ezawa1}.
After performing a partial gauge fixing, which leaves an abelian
gauge degree of freedom, one knows that point-like 
singularities in the three-dimensional space ${\bf R}^3$, 
associated with the homotopy group 
$\pi_{2}({\rm SU}(N_{c})/{\rm U}(1)^{N_{c}-1})=Z_{\infty}^{N_{c}-1}$,
can be identified as magnetic monopoles (QCD-monopoles) 
\cite{Hooft2,Ezawa1}.
If monopoles are condensed in the QCD vacuum, the dual Meissner effect
results \cite{Hooft3}.
Recent lattice QCD simulations actually show that the dual Meissner effect 
is brought to the true vacuum by QCD-monopole 
condensation\footnote[2]{QCD-monopole condensation is characterized by 
the presence of the long and tangled monopole trajectories in the 
four-dimensional space ${\bf R}^4$ and can be interpreted as 
the Kosterlitz-Thouless type phase transition.}
(see, \eg a recent review article \cite{Polikarpov}).
Hence, color confinement can be regarded 
as the dual version of superconductivity. 

It seems that instantons and QCD-monopoles are relevant topological objects 
for the description of distinct phenomena.
Here, we should emphasize that QCD-monopoles would play an essential role on
non-perturbative features of QCD, which include not only
confinement and but also S$\chi$SB.
This possibility was first studied by using the Schwinger-Dyson equation
with the gluon propagator in the background of 
condensed monopoles \cite{Sasaki1}.
The idea of providing S$\chi$SB due to 
QCD-monopole condensation was supported by
lattice simulations \cite{Miyamura1,Woloshyn}.
Thus, these results shed new light on the non-trivial 
relation between instantons and QCD-monopoles. 
Recently, both analytic works \cite{Brower,Suganuma1} and
numerical works 
\cite{Miyamura2,Markum,Teper,Suganuma2,Bornyakov} 
have shown the existence of the strong correlation 
between these topological objects in spite of the fact that 
they originate from different homotopy groups. 
Furthermore, monopole trajectories become more complicated
with the instanton density increasing in the background of a random 
ensemble of instanton solutions \cite{Fukushima}. 
These results seems to suggest that the instanton liquid and QCD-monopole 
condensation may be indistinguishable. 

In this paper, we would like to know whether 
the correlation between instantons and QCD-monopoles
has some physical significance, or not, from the topological viewpoint.
Here, we must not forget the hypothesis of abelian dominance, which 
Ezawa and Iwazaki had first advocated \cite{Ezawa1}. 
This hypothesis is essentially composed of two sentences \cite{Ezawa1}:
\begin{itemize}
   \item Only the abelian component 
         of gauge fields is relevant at a long-distance scale. 

   \item The non-abelian effects are mostly inherited by magnetic monopoles. 
\end{itemize}
Actually, lattice Monte Carlo simulation indicates that  
abelian dominance for some physical quantities, \eg the string tension 
\cite{Suzuki,Bari}, the chiral condensate \cite{Miyamura1,Woloshyn}
and also several light hadron spectra \cite{Miyamura3,Kitahara} is realized 
in the maximally abelian (MA) gauge as well as monopole dominance. 
In this gauge, at least, the abelian component of the gauge field could be 
an important dynamical degree of freedom at a long-distance scale. 
Here, an unavoidable question arises relating to the non-trivial correlation 
between instantons and QCD-monopoles. {\it In the abelian dominated system,
is it possible for the non-abelian topological nature to survive?}
For such an essential question, Ezawa and Iwazaki have proposed a 
remarkable conjecture \cite{Ezawa2}: 
once abelian dominance is postulated, the 
topological feature is preserved by the presence of monopoles.
The main purpose of this paper is to confirm the presence of
the non-trivial correlation between instantons and QCD-monopoles 
by finding evidence for the Ezawa-Iwazaki conjecture.

For simplicity, we restrict the
discussion to the case of SU(2) gauge group throughout this paper.
The organization of our paper is as follows. 
In Sec.2, we show that the topological charge is explicitly 
represented in terms of the monopole current and the abelian 
component of gauge fields through the hypothesis of abelian dominance. 
In Sec.3, we confirm numerically its justification by 
means of Monte Carlo simulation within SU(2) gauge theory
after the MA gauge fixing. Finally, Sec.4 is devoted to a summary and 
conclusions.
\section{Abelian dominance hypothesis for topological charge}
We first address the definition of the topological charge 
in the lattice formulation. 
The naive and field-theoretical definition of the topological 
density \cite{Ilgenfritz} is given by 
%
%
\beqn
q(s) \equiv \frac{1}{2} \epsilon_{\mu \nu \rho \sigma} \tr
\left \{ P_{\mu \nu}(s) P_{\rho \sigma}(s) \right \}\;,
\label{Eq:su2top}
\eeqn
with the clover averaged SU(2) plaquette:
%
%
\beqn
P_{\mu \nu}(s)\equiv\frac{1}{4}\left(\right.
U_{\mu \nu}(s) + U^{\dag}_{-\mu \nu}(s) 
+ U^{\dag}_{\mu -\nu}(s) + U_{-\mu -\nu}(s)
\left.\right)\;.
\eeqn
Here we have used the convenient notation for the SU(2) link variable;
$U_{-\mu}(s) = U^{\dag}_{\mu}(s-{\hat \mu})$
so that $U_{\pm\mu \pm\nu}(s)=U_{\pm\mu}(s)U_{\pm\nu}(s+{\hat \mu})
U^{\dag}_{\pm\mu}(s+{\hat \nu})U^{\dag}_{\pm\mu}(s)$.
One naively expects that the topological charge $Q_{\rm cont}$ 
is extracted from the summation of the previously defined topological 
density over all sites up to leading order in powers of the lattice spacing 
$a$ \cite{Ilgenfritz}:
%
%
\beqn
Q_{L}=-\frac{1}{16\pi^2}\sum_{s}q(s) \simeq Q_{\rm cont} + {\cal 
O}(a^6)\;,
\label{Eq:topchrL}
\eeqn
where $Q_{\rm cont}
=\frac{1}{16\pi^2}\sum_{s}\tr\left\{ a^{4} g^{2} G_{\mu \nu}(s) 
{}^{*}G_{\mu \nu}(s)\right\}$.
Strictly speaking, the value $Q_{L}$ has not only ${\cal O}(a^{6})$ corrections, 
but also a renormalized multiplicative correction of $Q_{\rm cont}$ in 
Eq.(\ref{Eq:topchrL}) \cite{Giacomo}. In general, one needs some smoothing 
method to eliminate undesirable ultraviolet fluctuations in the numerical 
simulation so that a renormalization factor would be brought to unity.

If we assume that the QCD vacuum is described as the 
abelian dominated system {\it in a suitable abelian gauge}, then
the SU(2) link variable is expected to be U(1)-like as 
$U_{\mu}(s) \simeq u_{\mu}(s) \equiv \exp \left\{
i\sigma_{3}\theta_{\mu}(s) \right\}$.
Here, we define the angular variable $\theta_{\mu}$ as
%
%
\beqn
\theta_{\mu}(s) \equiv {\rm arctan}[{U^{3}_{\mu}(s)}/{U^{0}_{\mu}(s)}]
\eeqn
in the compact domain $[-\pi, \pi)$ when the SU(2) link variable is 
parameterized as $U_{\mu}(s)=U^{0}_{\mu}(s)+i\sigma_{a}U^{a}_{\mu}(s)$ 
$(a=1,2,3)$ \cite{Kronfeld}. 
In the abelian dominated system, we might consider 
the abelian analog of the topological density 
$q_{_{\rm Abel}}(s)$ \cite{Bornyakov},
instead of the previously defined topological density,
through the replacement of $P_{\mu \nu}$ by the clover averaged 
U(1) plaquette $p_{\mu \nu}$:
%
%
\barr
p_{\mu \nu}(s) &\equiv& 
\frac{1}{4}\left(\right.
u_{\mu \nu}(s) + u^{\dag}_{-\mu \nu}(s) 
+ u^{\dag}_{\mu -\nu}(s) + u_{-\mu -\nu}(s)
\left.\right) \nonumber \\
&=&\frac{1}{4}\sum_{i,j=0}^{1}u_{\mu \nu}
(s-i{\hat \mu}-j{\hat \nu})\;,
\earr
where $u_{\mu \nu}$ denotes the U(1) elementary plaquette.
The explicit expression of $q_{_{\rm Abel}}(s)$
is then given by
%
%
\barr
q_{_{\rm Abel}}(s) &=& \frac{1}{2} \varepsilon_{\mu 
\nu \rho \sigma} {\rm tr}\left\{ p_{\mu \nu}(s) p_{\rho 
\sigma}(s) \right\} \nonumber \\
&=& - \frac{1}{16} 
\sum_{i,j,k,l=0}^{1}
\varepsilon_{\mu \nu \rho \sigma} 
\sin \theta_{\mu \nu}
(s-i{\hat \mu}-j{\hat \nu})
\sin \theta_{\rho \sigma}
(s-k{\hat \rho}-l{\hat \sigma})\;,
\earr
where $\theta_{\mu \nu}(s)\equiv 
\del_{\mu}\theta_{\nu}(s)-\del_{\nu}\theta_{\mu}(s)$ \cite{Sasaki2}.
$\del_{\mu}$ denotes the nearest-neighbor forward difference 
operator satisfying $\del_{\mu}f(s)\equiv f(s+{\hat \mu})-f(s)$.

Our next aim is to discuss the expression of the abelian analog of the
topological density 
in the naive continuum limit $a \rightarrow 0$.  
This is because we need only the leading order term in powers 
of the lattice spacing to determine the corresponding topological charge.
Here, one may notice that the U(1) elementary plaquette $u_{\mu \nu}$ 
is a multiple valued function of the U(1) plaquette angle $\theta_{\mu \nu}$. 
Hence, we divide $\theta_{\mu \nu}$ into two parts,
%
%
\beqn
\theta_{\mu \nu}(s) = {\bar \theta}_{\mu \nu}(s) + 2\pi n_{\mu 
\nu}(s)\;,
\eeqn
where ${\bar \theta}_{\mu \nu}$ is defined in the principal domain 
$[-\pi, \pi)$, which corresponds to the U(1) field strength. 
The anti-symmetric tensor $n_{\mu \nu}$ can take the restricted 
integer values $0,\pm 1,\pm 2$. 
Taking the limit $a \rightarrow 0$, \ie ${\bar \theta}_{\mu \nu} 
\rightarrow 0$, we thus arrive at the following expression 
\cite{Sasaki2}:
%
%
\beqn
q_{_{\rm Abel}}(s) \approx - \varepsilon_{\mu \nu \rho \sigma}
{\bar \Theta}_{\mu \nu}(s){\bar \Theta}_{\rho \sigma}(s)\;,
\label{Eq:abeltop}
\eeqn
where ${\bar \Theta}_{\mu \nu}(s) \equiv \frac{1}{4}\sum_{i,j=0}^{1}{\bar 
\theta}_{\mu \nu}(s-i{\hat \mu}-j{\hat \nu})$. 
The r.h.s of Eq.(\ref{Eq:abeltop}) does not
necessarily vanish in spite of the fact that it is represented only 
in terms of the U(1) field strength.

Next, we show that the non-vanishing contribution to the value of 
$q_{_{\rm Abel}}(s)$ results from monopoles.
For the identification of monopoles, we follow
DeGrand-Toussaint's definition in the compact U(1) gauge theory 
\cite{DeGrand}. The magnetic current is given by
%
%
\barr
k_{\mu}(s)
&\equiv& \frac{1}{4\pi}\epsilon_{\mu \nu \rho \sigma}\del_{\nu}
{\bar \theta}_{\rho \sigma}(s+{\hat \mu}) \nonumber \\
&=&-\frac{1}{2}\epsilon_{\mu \nu \rho \sigma}\del_{\nu}n_{\rho 
\sigma}(n+{\hat \mu})\;.
\earr
In the last line, we have used the Bianchi identity on the U(1) plaquette angle; 
$\epsilon_{\mu \nu \rho \sigma}\del_{\nu}\theta_{\rho \sigma}=0$.
Then we can easily see that the current $k_{\mu}$ denotes
the integer-valued magnetic current, \ie the 
monopole current \cite{DeGrand}. 
Strictly speaking, the monopole current resides 
at the dual lattice site orthogonal to direction $\mu$.
For simplicity, we have used an undiscriminating notation between the ordinary 
lattice site and the dual lattice site. One may notice that it is obvious that $k_{\mu}$ is topologically 
conserved; $\del^{\prime}_{\mu}k_{\mu}=0$.
$\del^{\prime}$ denotes the nearest-neighbor backward difference 
operator satisfying $\del^{\prime}_{\mu}f(s)\equiv f(s)-f(s-{\hat \mu})$.

Here, we define the averaged magnetic current\footnote[2]{In some 
sense, this is similar to the type-II extended monopole current 
\cite{Ivanenko}.} 
in the $2^3$ extended cube at the center of the lattice site $s$ orthogonal 
to direction $\mu$ \cite{Sasaki2} as
%
%
\beqn
{\cal K}_{\mu}(s)
\equiv \frac{1}{8}\sum_{i,j,k=0}^{1}k_{\mu}(s-i{\hat \nu}-j{\hat 
\rho}-k{\hat \sigma}) \;,
\eeqn
where there indices $(\nu, \rho, \sigma)$ are complementary to $\mu$.
Using the nearest-neighbor central difference operator; 
$\Delta_{\mu}\equiv\frac{1}{2}\{\del_{\mu}+\del^{\prime}_{\mu}\}$,
we therefore rewrite ${\cal K}_{\mu}$ as the following form:
%
%
\beqn
{\cal K}_{\mu}(s)
=\frac{1}{4\pi}\epsilon_{\mu \nu \rho 
\sigma}\Delta_{\nu}{\bar \Theta}_{\rho \sigma}(s)\;,
\eeqn
which obviously satisfies the 
conservation law; $\Delta_{\mu}{\cal K}_{\mu}(s)=0$.

To show the explicit contribution of monopoles to the topological
charge, we introduce the dual potential 
${\cal B}_{\mu}$ satisfying the following equation \cite{Smit}:
%
%
\beqn
\left(\Delta^2\delta_{\mu \nu}
-\Delta_{\mu}\Delta_{\nu} \right){\cal B}_{\nu}(s)
=-2\pi{\cal K}_{\mu}(s)\;.
\eeqn
where $\Delta^2 \equiv \Delta_{\mu}\Delta_{\mu}$.
We can perform the Hodge decomposition on the clover-average U(1) 
field strength ${\bar \Theta}_{\mu \nu}$
with the dual potential ${\cal B}_{\mu}$ as
%
%
\beqn
{\bar \Theta}_{\mu \nu}(s)=\Delta_{\mu}{\cal A}_{\nu}(s)
-\Delta_{\nu}{\cal A}_{\mu}(s)
+\varepsilon_{\mu \nu \rho 
\sigma}\Delta_{\rho}{\cal B}_{\sigma}(s)\;,
\eeqn
where ${\cal A}_{\mu}$ is the Gaussian fluctuation 
\cite{Smit}.
After a little algebra using the partial summation,
we find the explicit contribution of monopoles to the
r.h.s of Eq.(\ref{Eq:abeltop}) \cite{Sasaki2} as 
%
%
\barr
\varepsilon_{\mu \nu \rho \sigma}
{\bar \Theta}_{\mu \nu}(s){\bar \Theta}_{\rho \sigma}(s)
&=&4\varepsilon_{\mu \nu \lambda \omega}\varepsilon_{\lambda \omega \rho \sigma}
\Delta_{\mu}{\cal A}_{\nu}(s)
\Delta_{\rho}{\cal B}_{\sigma}(s) \nonumber \\
&=&16\pi {\cal A}_{\mu}(s) {\cal K}_{\mu}(s) + 
\cdot\cdot\cdot\;,
\earr
where the ellipsis stands for the total divergence, which will drop 
in the summation over all sites.
Consequently, we arrive at the conjecture that 
{\it the topological feature is preserved by 
the presence of monopoles in the abelian dominated system} 
\cite{Sasaki2}:
%
%
\beqn
Q_{\rm cont}\simeq - \frac{1}{16\pi^2}\sum_{s}q_{_{\rm Mono}}(s)\;,
\eeqn
where $q_{_{\rm Mono}}(s)\equiv -16\pi {\cal A}_{\mu}(s)
{\cal K}_{\mu}(s)$. In addition, the Gaussian fluctuation can not be definitely 
identified except for the Landau gauge solution;
$\Delta_{\mu}{\cal A}^{L}_{\mu}(s)=0$ where a superscript $L$ denotes `Landau'.
However, we need only to know the Gaussian fluctuation in the Landau gauge on the 
measurement of the quantity $\sum_{s}q_{_{\rm Mono}}(s)$, which has the
residual U(1) gauge invariance.
\section{Numerical results}
This section is devoted to a numerical analysis to justify the
above conjecture through Monte Carlo simulation within 
SU(2) lattice gauge theory using the standard Wilson action.
We choose the MA gauge as an applicable abelian gauge for the realization 
of abelian dominance in this paper. As we pointed out, 
the QCD vacuum has been observed as an abelian 
dominated system by lattice Monte Carlo simulations \cite{Polikarpov}.
We shall return to the explicit procedure of the MA gauge fixing later.

As we have mentioned before, one needs to smooth the Monte Carlo gauge 
configurations in order to determine the topological charge.
To eliminate undesirable fluctuations, we adopt the 
naive cooling method which is an iterative scheme to replace 
each link variable using the following procedure \cite{Ilgenfritz}:
%
%
\beqn
U_{\mu}(s) \rightarrow U^{\prime}_{\mu}(s) = \Sigma_{\mu}(s)
/ ||\Sigma_{\mu}(s)|| \;\;,
\eeqn
where $||\Sigma_{\mu}(s)||$ denotes the square root of 
determinant of $\Sigma_{\mu}$ 
and
%
%
\beqn
\Sigma_{\mu}(s)=\sum_{\nu\neq\mu}\left(
U_{\nu}(s)U_{\mu}(s+{\hat \nu})U^{\dag}_{\nu}(s+{\hat \mu})
+U^{\dag}_{\nu}(s-{\hat \nu})U_{\mu}(s-{\hat \nu})
U_{\nu}(s+{\hat \mu}-{\hat \nu})
\right) \;.
\eeqn
This procedure is derived from the condition to minimize the action
through the variation of the link variable \cite{Ilgenfritz}. Then, an
iterated process leads to a local minimum of the 
action, which corresponds to a solution to the field equation. The 
relation
$Q_{L}\approx Q_{\rm cont}$ is expected through the cooling procedure.

Before turning to the abelian gauge fixing, we must take account of 
the permutation between the cooling procedure and the gauge fixing procedure.
The action in the compact lattice gauge theory 
is not usually affected by the gauge fixing procedure for the link variables 
because of the gauge-invariant measure. However, the fundamental
modular region \cite{Zwanziger} appears in the case of the 
MA gauge fixing \cite{Chernodub}. 
Then, the action is modified by the contribution 
from the Faddeev-Popov determinant \cite{Chernodub}.
In other words, the above cooling procedure is not justified 
after the MA gauge fixing.
Thus, we have to apply the MA gauge fixing procedure to
the resulting gauge configuration after several cooling sweeps.

In the lattice formulation, the MA gauge fixing is defined by 
maximizing the gauge dependent variable $R$
under gauge transformations; $U_{\mu}(s) \rightarrow {\tilde U}_{\mu}
(s)=\Omega(s)U_{\mu}(s)\Omega^{\dag}(s+{\hat \mu})$ \cite{Kronfeld}:
%
%
\beqn
R[\Omega]
=\sum_{s,\;\mu}\tr\left\{
\sigma_{3} {\tilde U}_{\mu}(s) \sigma_{3} {\tilde U}^{\dag}_{\mu}(s) 
\right\}
=2\sum_{s,\;\mu}\left( 
1-2\chi_{\mu}(s) 
\right) \;,
\label{magauge}
\eeqn
where $\chi_{\mu}\equiv
({\tilde U}^{1}_{\mu}(s))^2+ ({\tilde U}^{2}_{\mu}(s))^2$.
The maximization of $R$ implies that the off-diagonal components of the
SU(2) link variables are minimized as much as possible
through the gauge transformation.

After the MA gauge fixing, we extract the U(1) field strength 
${\bar \theta}_{\mu \nu}$ and the monopole current $k_{\mu}$ 
from the ${\rm U}(1)$ link variable following the previous section.  
The Gaussian fluctuation in the Landau gauge is computed by the convolution of 
the electric current with the lattice Coulomb propagator 
${\cal D}(s-s^{\prime})$ 
%
%
\beqn
{\cal A}^{L}_{\mu}(s)=-\sum_{s^{\prime}}{\cal D}(s-s^{\prime})\Delta_{\lambda}{\bar 
\Theta}_{\lambda \mu}(s^{\prime})\;,
\eeqn
where the lattice Coulomb propagator 
satisfies the equation; $\Delta^{2}{\cal D}(s-s^{\prime})=-\delta_{s,\;s^{\prime}}$.
One can obtain the explicit form of ${\cal D}(s-s^{\prime})$ on a $L^{4}$ lattice as
%
%
\beqn
{\cal D}(s-s^{\prime})=\sum_{p}{1 \over {\sum_{\mu}
\sin^2(p_{\mu})}}e^{ip\cdot (s-s^{\prime}) }
\label{Eq:cgreen}
\eeqn
with the following abbreviation: $p_{\mu}\equiv \frac{2\pi}{L}n_{\mu}$ and 
$\sum_{p}\equiv \frac{1}{L^{4}} \Pi_{\mu}\sum_{n_{\mu}=0}^{L-1}$.

We generate the gauge configurations by using the 
heat bath algorithm on a $16^4$ lattice at $\beta=2.4$.
All measurements have been performed on independent configurations, which 
are separated from each other by 500 sweeps after the system has reached thermal
equilibrium for 1000 heat bath sweeps.
For the MA gauge fixing in the numerical simulation, we use the overrelaxation 
method with parameter $\omega =1.7$ \cite{Mandula}. 
Let us concentrate on the realization of the abelian dominance 
for topological density in the MA gauge. 
Before turning to the topological density, we show 
the probability distribution of the amplitude of the off-diagonal 
component $\chi_{\mu}$ in Fig.\ref{fig:chi}. 
The solid curve, the broken curve and the dotted curve denote the case 
for configurations before cooling and after cooling for 1 sweep and 
30 sweeps, respectively.
This figure tells us that the off diagonal components of the SU(2) link
variable are actually made as small as possible on the whole 
lattice in the MA gauge. Furthermore, such a tendency becomes prominent 
with more cooling sweeps.
The information from Fig.\ref{fig:chi} suggests that
large numbers of cooling sweeps could enhance abelian dominance. 
In addition, the off diagonal components; $U_{\mu}^1$ and $U_{\mu}^2$
behave like random variables on the whole lattice
before the MA gauge fixing so that the 
probability distribution is flat on unity in the case of no gauge 
fixing whether configurations are cooled, or not.

We have checked the above expectation on the measurement of the topological 
density $q(s)$ and the abelian analog of topological density $q_{\rm Abel}(s)$.
Fig.2 and Fig.3 display $q(s)$ and $q_{\rm Abel}(s)$ in a two-dimensional 
plane after 30 and 100 cooling sweeps.
We define this plane as a slice at the center of 
the instanton in the $(z,t)$-plane.
The center is identified as the local maximum of $q(s)$.
The topological density around the instanton is little affected by the
cooling procedure as is shown in Fig.\ref{fig:full30} and
Fig.\ref{fig:full100}. For the abelian analog of topological density, 
such a stability in the cooling process is not required because of the lack 
of the topological basis in the U(1) manifold.
At 30 cooling sweeps, the abelian dominance for topological
density is not largely revealed as shown in Fig.\ref{fig:abel30}, 
since a large amplitude of the off-diagonal components still remains locally, 
especially around the center of the instanton and/or the monopole.
However, after 100 cooling sweeps,
one can recognize that a similar lump to the topological density
is located around the center of the instanton in Fig.\ref{fig:abel100}.
Thus, the topological density is dominated by the abelian 
analog of topological density after enough cooling sweeps, as expected.

Next, we will examine the correlation between 
topological charge and monopoles.
The two types of corresponding topological charge are computed through
the following formula:
%
%
\barr
Q_{\rm SU(2)}&\equiv& - \frac{1}{16 \pi^2} \sum_{s} q(s) \;,\\
Q_{\rm Mono}&\equiv& - \frac{1}{16 \pi^2} \sum_{s} q_{_{\rm Mono}}(s)\;.
\earr
$Q_{\rm SU(2)}$ denotes the ordinary topological charge.
$Q_{\rm Mono}$ denotes the corresponding topological charge via 
monopoles.
We show the scatter plots of $Q_{\rm SU(2)}$ vs. $Q_{\rm Mono}$
in Fig.4; (a) no gauge fixing and (b) the MA gauge fixing 
after 100 cooling sweeps for 50 independent configurations. 
Obviously, there is not any correlation between the two topological charges 
{\it before the MA gauge fixing}.
A one-to-one correspondence between $Q_{\rm SU(2)}$ 
and $Q_{\rm Mono}$ is revealed in the scatter plot {\it after the MA gauge fixing}.
In further detail, we find that there is a small variance between 
$Q_{\rm SU(2)}$ and $Q_{\rm Mono}$. 
The slope of the linear correlation 
in the scatter plot is not unity, but 1.43 in Fig.\ref{fig:mag}
($16^4$ at $\beta=2.4$). 
On the several measurements at $\beta=2.45$ and $2.5$, it
seems that this slope hardly depends on the lattice spacing after 
large enough numbers of cooling sweeps (see Table 1).
Table 1 tells us that the relation;
$Q_{\rm Mono}\approx 0.7 Q_{\rm SU(2)}$ is almost satisfied in the 
MA gauge on these data.

%
%
\begin{center}

\leavevmode
\hbox{
     \begin{tabular}[t]{c|ccc}
     \hline\hline
     & \multicolumn{3}{|c}{cooling sweeps} \\
     \cline{2-4} 
	 \makebox[2cm]{$\beta$}  
	 & \makebox[1.8cm]{30} & \makebox[1.8cm]{50}& \makebox[1.8cm]{100} \\
     \hline
     2.40         & 1.45 & 1.46 & 1.43 \\
     2.45         & 1.46 & 1.43 & 1.44 \\
     2.50         & 1.49 & 1.43 & 1.43 \\
     \hline\hline
     \end{tabular}
}
\end{center}
\hspace{0.5cm}

To discuss the topological feature on $Q_{\rm Mono}$, we show the 
probability distribution of $Q_{\rm Mono}$ in Fig.\ref{fig:dis} by using 3000 
independent configurations at $\beta=2.4$ on a $16^4$ lattice.
Several dotted lines correspond to partial contributions to the whole 
distribution, which are assigned to some integer value of $Q_{\rm SU(2)}$.
We find several discrete peaks in Fig.\ref{fig:dis}, since
each partial contribution is a Gaussian-type 
distribution with the half width slightly less than unity 
around discrete values ($\approx 0.7 Q_{\rm SU(2)}$).
This implies that $Q_{\rm Mono}$ is classified by 
approximately discrete values. 
Namely, it seems that monopoles almost inherit the topological nature
of the original gauge configurations.

\section{Summary and Conclusion}

Our work was motivated by the Ezawa-Iwazaki conjecture:
the topological feature is preserved by the presence of QCD-monopoles 
once abelian dominance is postulated.
To confirm this, we have performed an analytical study of 
how the monopole current is related to the topological charge within 
SU(2) lattice gauge theory.
We consider the abelian analog of topological density, which is 
defined through the replacement of the SU(2) link variable by the 
U(1)-like link variable. Although this quantity is represented by 
the corresponding U(1) field strength in the naive continuum limit, 
it can carry the non-trivial contribution to the topological charge 
because of the presence of QCD-monopoles.
If we assume that the abelian component of gauge fields plays an essential 
role in the description of the long range physics, QCD-monopoles are 
required to inherit the topological feature.

Monte Carlo simulations have brought us
encouraging results for the above conjecture.
We have chosen the MA gauge as an applicable abelian gauge.
We have measured both the ordinary topological 
charge $Q_{\rm SU(2)}$ and the corresponding topological charge via 
monopoles, $Q_{\rm Mono}$.
In spite of the fact that there is no correlation between $Q_{\rm 
SU(2)}$ and $Q_{\rm Mono}$ without abelian 
gauge fixing, we found a one-to-one correspondence between the two charges 
in the MA gauge. And also the relation: $Q_{\rm Mono} \approx 0.7 Q_{\rm SU(2)}$
can be read off from our resulting data.   
Furthermore, $Q_{\rm Mono}$ is classified by approximately discrete 
values. Thus we have concluded that the topological nature is substantially 
inherited by QCD-monopoles. 
This conclusion is consistent with our previous study \cite{Sasaki3}, 
which shows that QCD-monopoles strongly correlate with the presence of 
fermionic zero-modes in the MA gauge.

\newpage
\section*{Acknowledgment}
We gratefully acknowledge useful discussions with H. Suganuma
and  Ph. de Forcrand.
We also thank T. Blum for encouragement and reading the manuscript.
All lattice simulations in this paper have been performed on DEC
Alpha Server 8400 5/440 at the Research Center for Nuclear Physics (RCNP)
of Osaka University. One of us (O.M.) appreciates support from
the Grant in Aid for Scientific Research by the Ministry of Education
(A) (No.0830424). Finally, another (S.S.) would like to thank 
all the members of Yukawa Institute for Theoretical Physics (YITP), 
especially, T. Matsui and K. Itakura for their warm hospitality during
his residence at the YITP of Kyoto University, where 
most of the present study has been carried out.

\newpage
\centerline{\large FIGURE CAPTIONS}
\begin{description}

\vspace{1.0cm}
\item[FIG.1.]
\begin{minipage}[t]{13cm}
\baselineskip=20pt
The probability distribution of the amplitude of the off-diagonal
components $\chi$ on a $16^{4}$ lattice at $\beta =2.4$.
\end{minipage}

\vspace{1.0cm}
\item[FIG.2.]
\begin{minipage}[t]{13cm}
\baselineskip=20pt
The topological density as a function of $z$ and $t$
in a two dimensional slice at the center of the instanton
after 30 cooling sweeps (a), and after 100 cooling sweeps (b) 
at $\beta=2.4$ on a $16^4$ lattice.
\end{minipage}

\vspace{1.0cm}
\item[FIG.3.]
\begin{minipage}[t]{13cm}
\baselineskip=20pt
The abelian analog of topological density as a function of $z$ and $t$
in a two dimensional slice at the center of the instanton
after 30 cooling sweeps (a), and after 100 cooling sweeps (b) 
at $\beta=2.4$ on a $16^4$ lattice.
\end{minipage}

\vspace{1.0cm}
\item[FIG.4.]
\begin{minipage}[t]{13cm}
\baselineskip=20pt
Scatter plot of $Q_{\rm SU(2)}$ vs. $Q_{\rm Mono}$ 
(a) before the MA gauge fixing  and (b) after the MA gauge fixing
at 100 cooling sweeps.
\end{minipage}

\vspace{1.0cm}
\item[FIG.5.]
\begin{minipage}[t]{13cm}
\baselineskip=20pt
Probability distribution of $Q_{\rm Mono}$ in the MA gauge fixing
at 100 cooling sweeps by using 3000 independent configurations.
\end{minipage}

\vspace{1.0cm}
\item[TABLE 1.]
\begin{minipage}[t]{13cm}
\baselineskip=20pt
Slope of the linear correlation in the scatter plot of
$Q_{\rm SU(2)}$ vs. $Q_{\rm Mono}$ on a $16^4$ lattice
at $\beta=2.4, 2.45$ and 2.5 after 30, 50 and 100 cooling sweeps.
\end{minipage}

\end{description}
\newpage
\centerline{\Large FIG.1 (Phys.Rev.D) Shoichi Sasaki \etal}

\vspace*{2.5cm}
%
%
\noindent
\centerline{\epsfxsize=15.0cm
\epsfbox{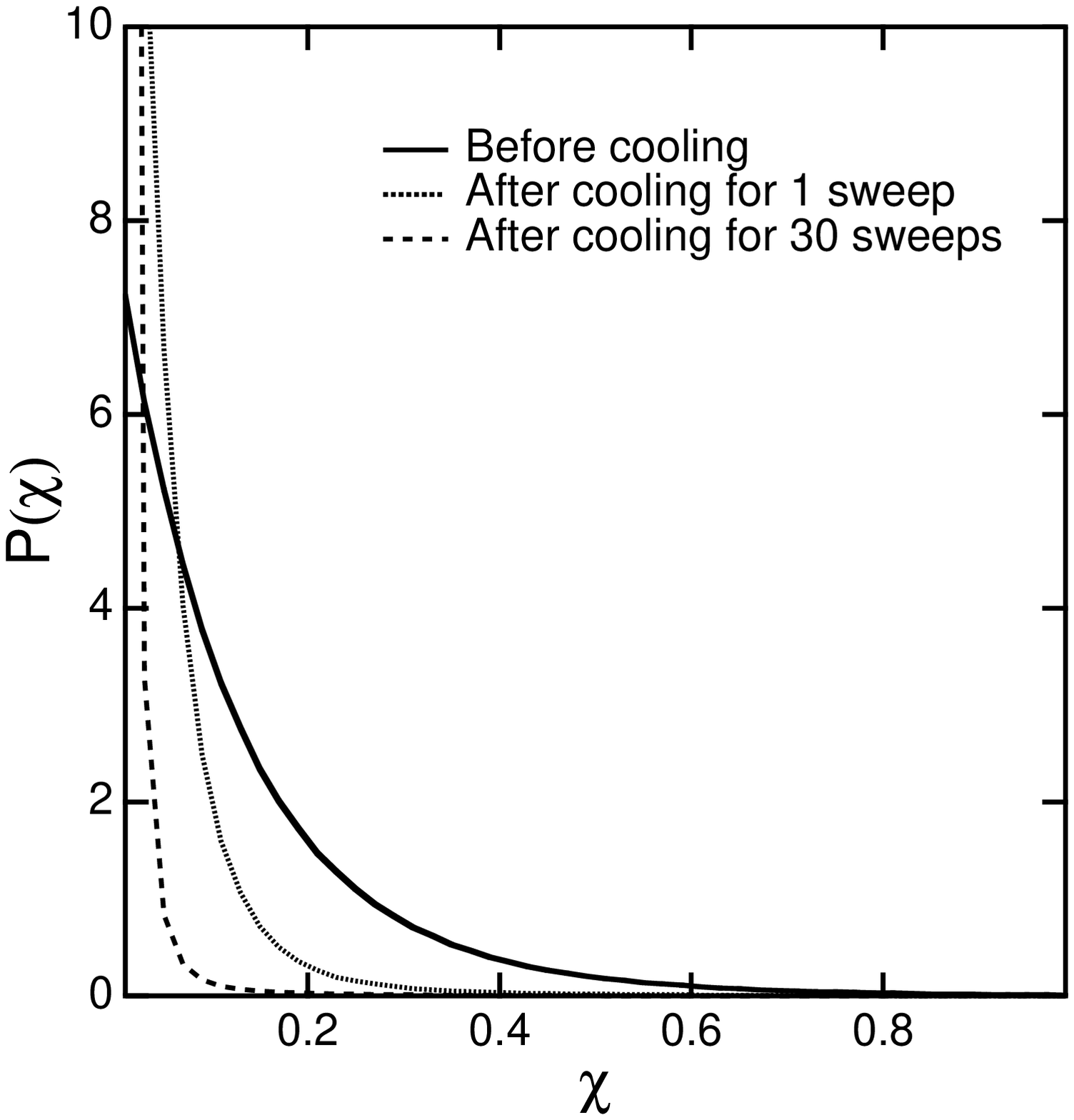}}
\vspace{0.5cm}
\centerline{\fcaption{\label{fig:chi}}}

\newpage
\centerline{\Large FIG.2a (Phys.Rev.D) Shoichi Sasaki \etal}

\vspace*{2.5cm}
{\setcounter{enumi}{\value{figure}}
\addtocounter{enumi}{1}
\setcounter{figure}{0}
\renewcommand{\thefigure}{\arabic{enumi}(\alph{figure})}

%
%
\noindent
\centerline{\epsfxsize=15.0cm
\epsfbox{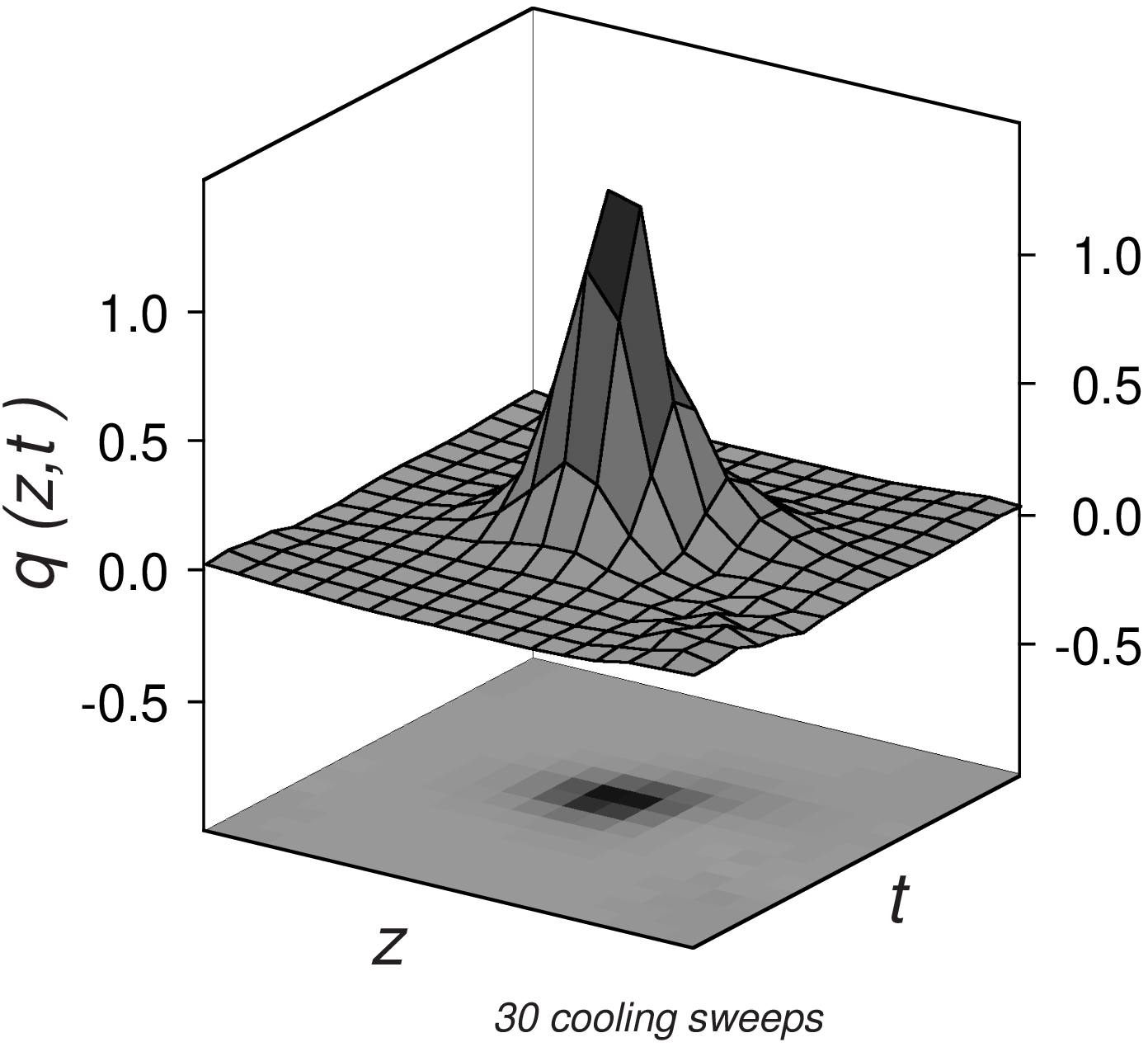}}
\vspace{0.5cm}
\centerline{\fcaption{\label{fig:full30}}}

\newpage
\centerline{\Large FIG.2b (Phys.Rev.D) Shoichi Sasaki \etal}

\vspace*{2.5cm}
\noindent
%
%
%
\noindent
\centerline{\epsfxsize=15.0cm
\epsfbox{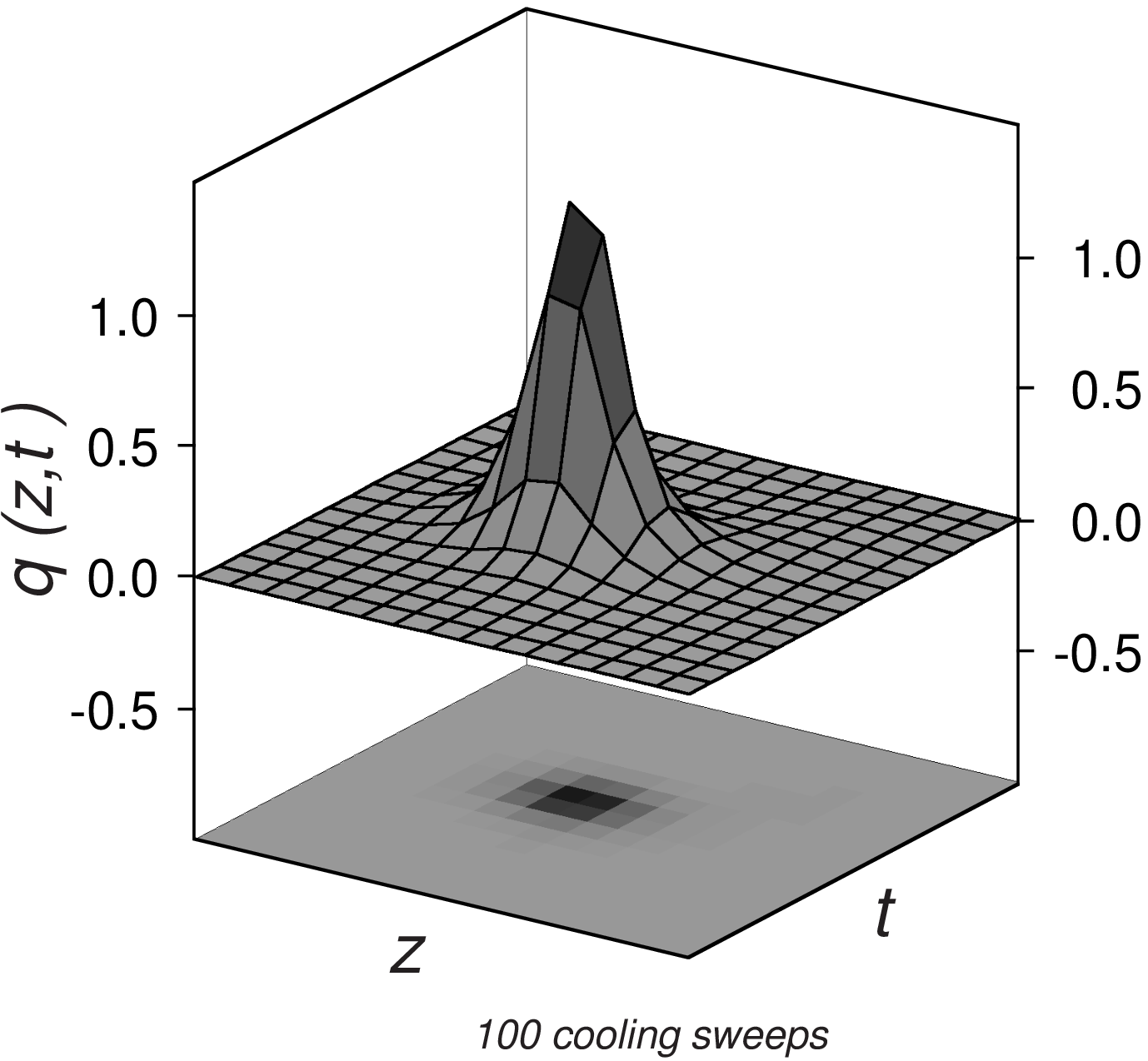}}
\vspace{0.5cm}
\centerline{\fcaption{\label{fig:full100}}}
\setcounter{figure}{\value{enumi}}
}

\newpage
\centerline{\Large FIG.3a (Phys.Rev.D) Shoichi Sasaki \etal}

{\setcounter{enumi}{\value{figure}}
\addtocounter{enumi}{1}
\setcounter{figure}{0}
\renewcommand{\thefigure}{\arabic{enumi}(\alph{figure})}

\vspace*{2.5cm}
%
%
\noindent
\centerline{\epsfxsize=15.0cm
\epsfbox{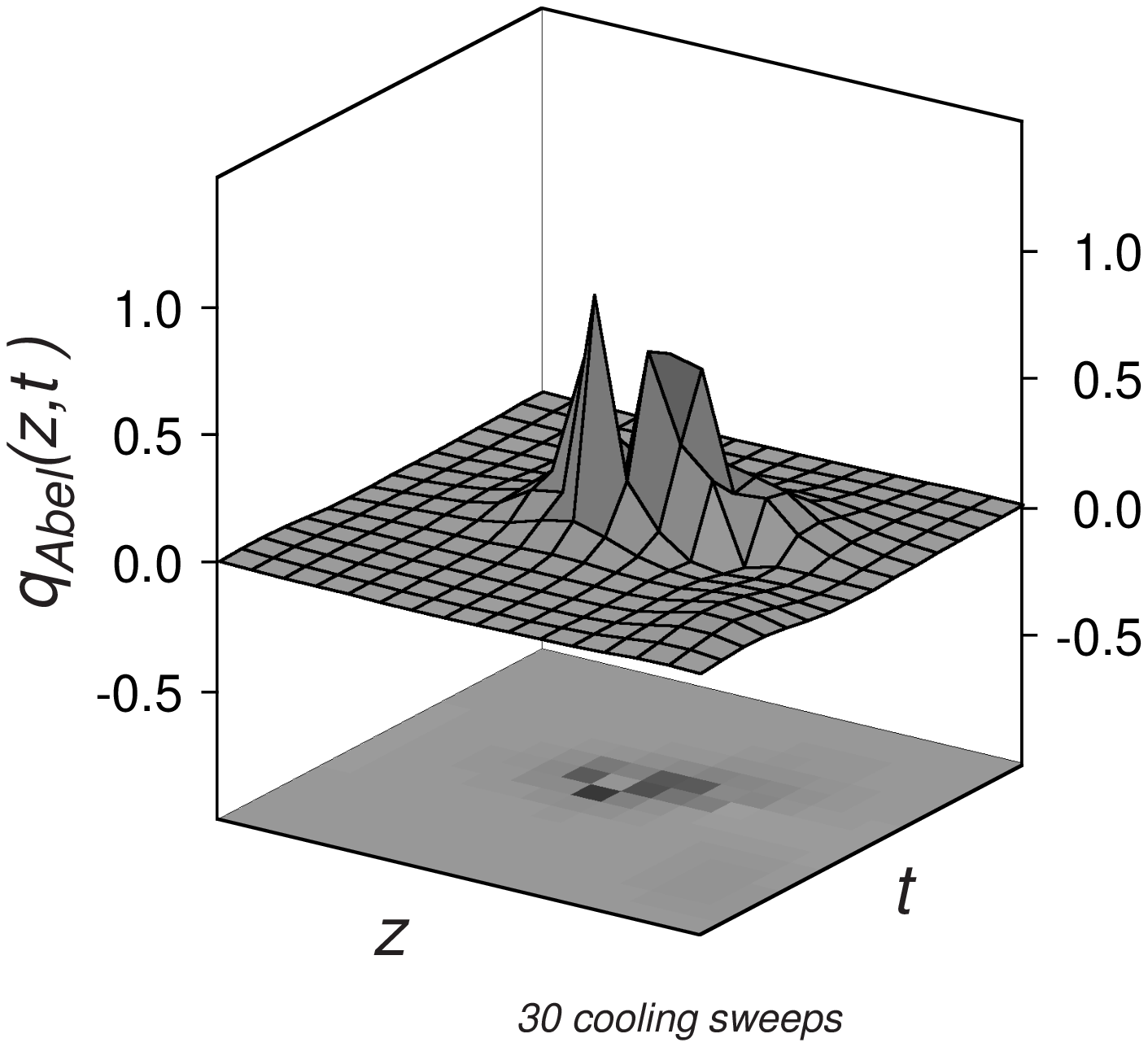}}
\vspace{0.5cm}
\centerline{\fcaption{\label{fig:abel30}}}

\newpage
\centerline{\Large FIG.3b (Phys.Rev.D) Shoichi Sasaki \etal}

\vspace*{2.5cm}
%
%
%
\noindent
\centerline{\epsfxsize=15.0cm
\epsfbox{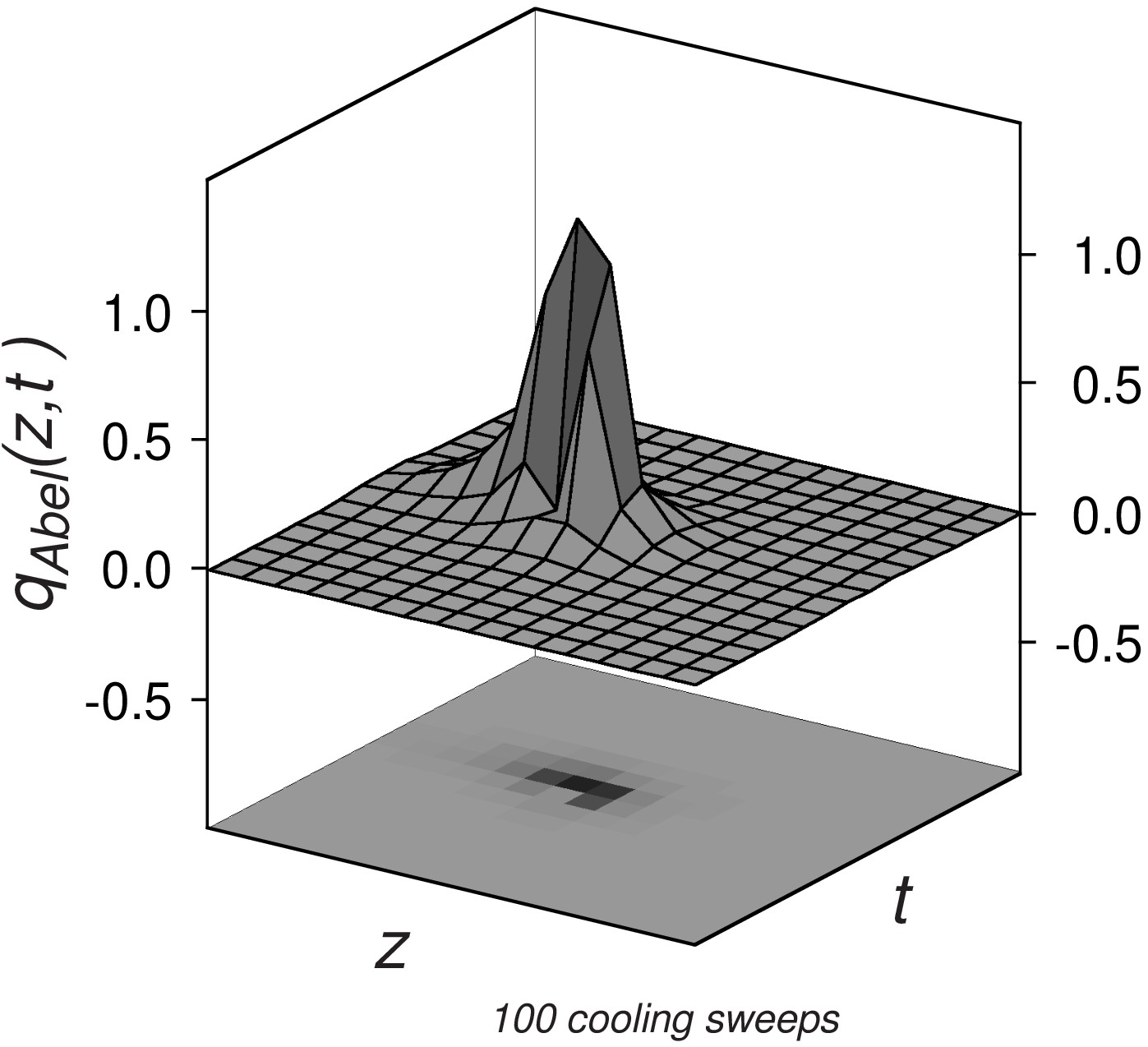}}
\vspace{0.5cm}
\centerline{\fcaption{\label{fig:abel100}}}
\setcounter{figure}{\value{enumi}}
}

\newpage
\centerline{\Large FIG.4a (Phys.Rev.D) Shoichi Sasaki \etal}

{\setcounter{enumi}{\value{figure}}
\addtocounter{enumi}{1}
\setcounter{figure}{0}
\renewcommand{\thefigure}{\arabic{enumi}(\alph{figure})}

\vspace*{2.5cm}
%
%
\noindent
\centerline{\epsfxsize=15.0cm
\epsfbox{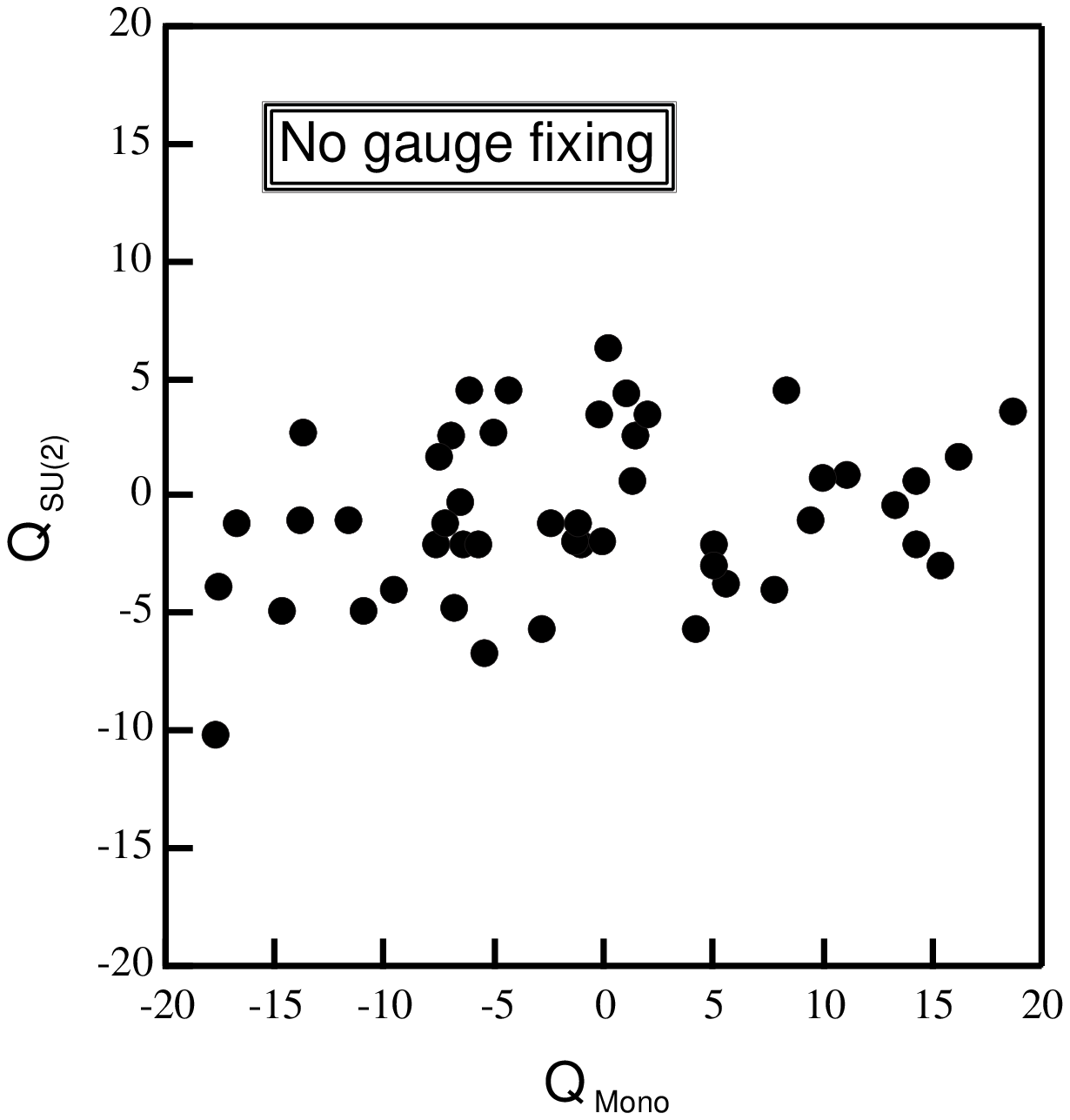}}
\vspace{0.5cm}
\centerline{\fcaption{\label{fig:ngf}}}

\newpage
\centerline{\Large FIG.4b (Phys.Rev.D) Shoichi Sasaki \etal}

\vspace*{2.5cm}
%
%
\noindent
\centerline{\epsfxsize=15.0cm
\epsfbox{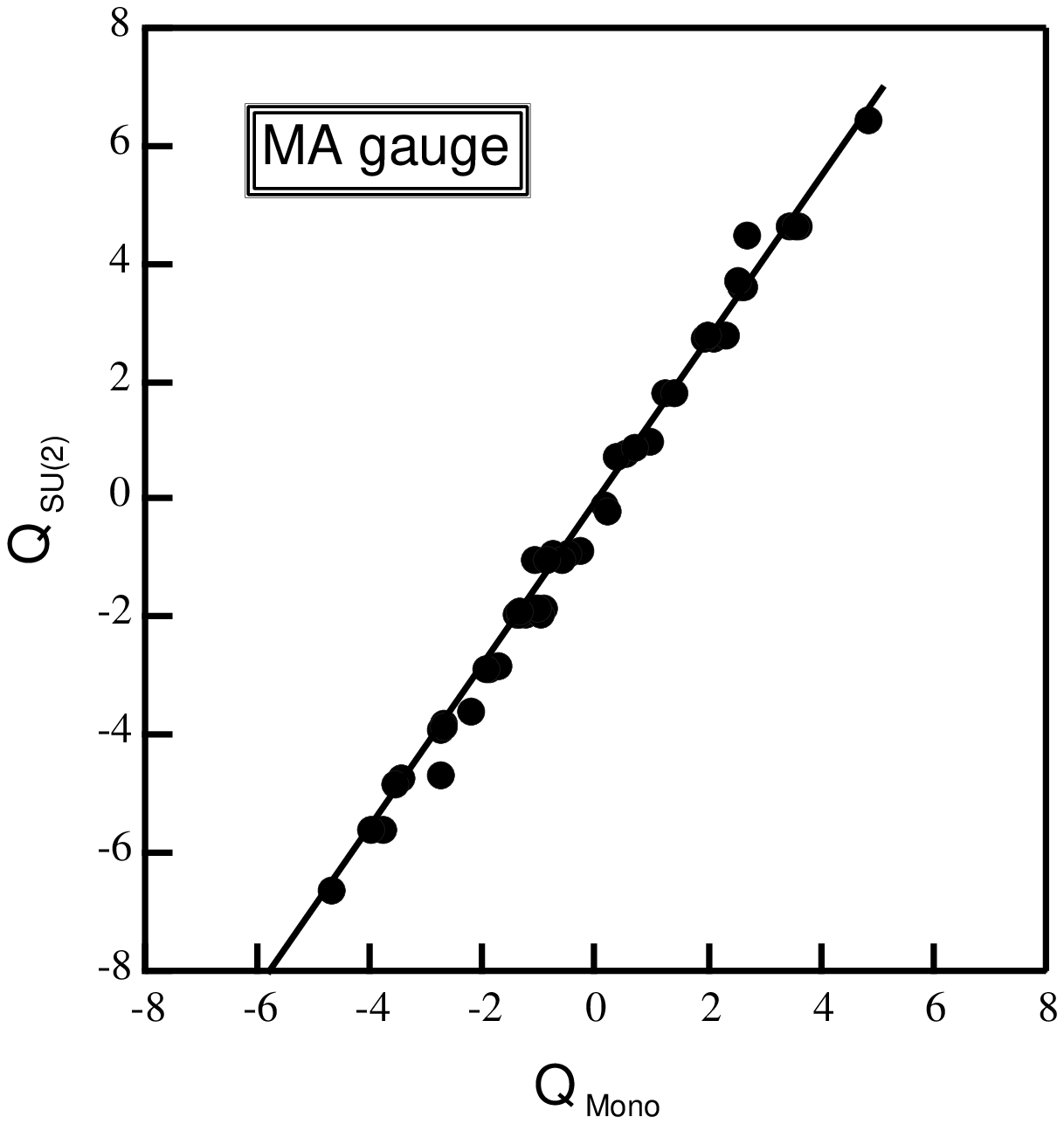}}
\vspace{0.5cm}
\centerline{\fcaption{\label{fig:mag}}}
\setcounter{figure}{\value{enumi}}
}

\newpage
\centerline{\Large FIG.5 (Phys.Rev.D) Shoichi Sasaki \etal}

\vspace*{2.5cm}
%
%
\noindent
\centerline{\epsfxsize=15.0cm
\epsfbox{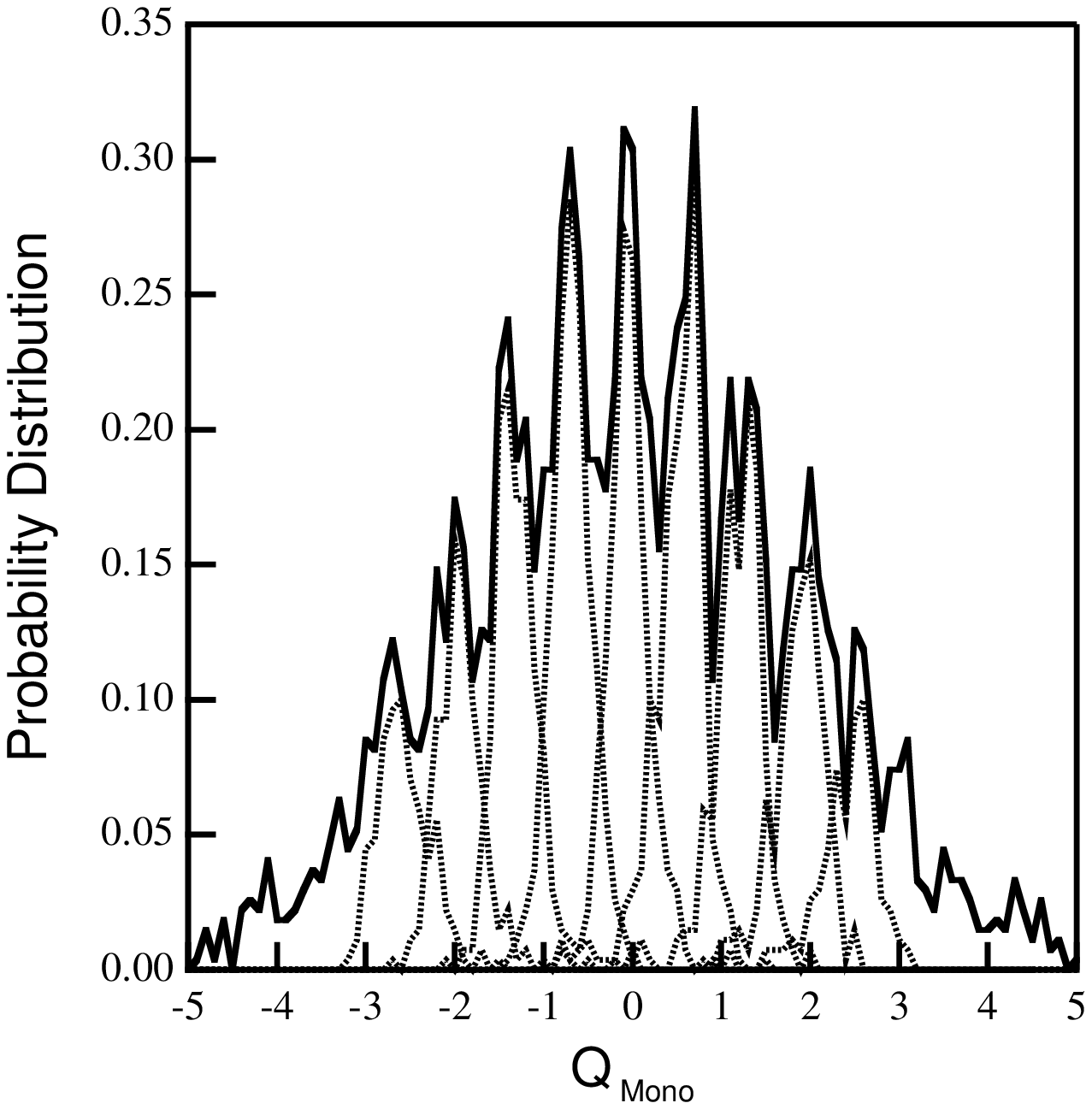}}
\vspace{0.5cm}
\centerline{\fcaption{\label{fig:dis}}}

%

\end{document}